# Adaptive self-organized criticality in deep neural networks

Simon Vock[1,2,3,4] and Christian Meisel[1,2,4,5]

[1] Section on Computational Neurology, Department of Neurology, Charité – Universitätsmedizin Berlin, Berlin, Germany
[2] Berlin Institute of Health, Berlin, Germany
[3] Faculty of Life Sciences, Humboldt University Berlin, Germany
[4] Bernstein Center for Computational Neuroscience, Berlin, Germany
[5] Center for Stroke Research, Berlin, Germany

## Abstract

Deep neural networks are high-dimensional dynamical systems whose function depends on the stable propagation of activity and perturbations across many layers. Maintaining suitable dynamical regimes may therefore be essential for robust learning and for preventing dynamical instabilities during training. Here, we show that the global dynamical state of a deep neural network can be autonomously regulated by purely local homeostatic plasticity. Neuronal activity is inferred from responses across inputs, and individual synapses are strengthened or weakened using only the activity of their postsynaptic neuron. Without measuring any global network property, this rule drives networks from both subcritical and supercritical initial conditions toward a common critical state, characterized by conserved activity propagation and a vanishing largest finite-time Lyapunov exponent. When combined with gradient-based learning, homeostatic adaptation counteracts the training-induced drift toward supercritical dynamics, while revealing a competition between dynamical regulation and task optimization. Our results demonstrate how adaptive self-organization can be implemented in deep neural networks and how local plasticity can control their collective dynamical operating point.

## 1. Introduction

Near continuous phase transitions, systems exhibit collective properties, including enhanced susceptibility, long-range correlations and a broad repertoire of dynamical responses that can benefit computation and memory[1–4]. Together with empirical signatures of criticality in neuronal activity, these properties have motivated the hypothesis that brain networks operate near critical states[5–8]. Self-organized criticality provides a mechanism by which such states can emerge without external tuning[9]. In adaptive networks, feedback between fast network dynamics and slower structural adaptation can drive this self-organization. Bornholdt and Rohlf showed that a simple local rule - where inactive nodes gain connections and active nodes lose them - drives global connectivity toward the critical point[10]. Similar principles extend to biologically more realistic neuronal networks, where activity-dependent synaptic plasticity can self-organize network dynamics toward criticality[11]. Thus, local dynamical information can be sufficient to regulate a global critical state.

The rapid advances in artificial intelligence (AI) have largely been driven by scaling deep neural networks (DNNs) - increasing model size, data, and computational resources[12,13]. Yet DNNs are also high-dimensional dynamical systems whose function depends on how activity propagates through successive layers. Weak effective coupling attenuates activity and perturbations, whereas strong coupling amplifies them; between these regimes lies a critical transition characterized by balanced signal propagation. In our previous work, we characterized this transition by a branching parameter close to unity, a maximum finite-time Lyapunov exponent close to zero, and maximal dynamic range[14]. We further showed that successful modern DNNs tend to operate close to this regime, that proximity to criticality is associated with improved task performance, and that explicitly maintaining critical dynamics can improve learning and mitigate loss of plasticity and model collapse. These findings identify criticality as a functionally relevant dynamical regime for deep learning.

Conventional mechanisms for controlling DNN dynamics generally rely on globally defined optimization objectives or externally chosen architectural and training parameters. Likewise, directly constraining criticality through an objective based on the input-output Jacobian requires explicit knowledge of a global property of the network. Adaptive self-organized criticality suggests a qualitatively different possibility: a network might regulate its global dynamical state through local plasticity alone, without measuring its distance from the critical point and without knowing the value of the corresponding global control parameter[15]. Here, we investigate whether such adaptive self-organization can occur in deep neural networks. Inspired by activity-dependent plasticity in adaptive network models, we introduce a simple homeostatic synaptic rule in which neurons with low activity strengthen incoming connections, whereas highly active neurons weaken them. We show that this purely local mechanism autonomously drives deep networks from both subcritical and supercritical initial conditions toward a common critical dynamical regime.

## 2. Model and adaptive self-organization

### A. Deep neural network dynamics

We consider feed-forward deep neural networks as discrete-time dynamical systems, where propagation from one layer to the next corresponds to successive steps of the dynamics. Unless stated otherwise, we use fully connected multilayer perceptrons with variable widths (N=1000, 1600) neurons per hidden layer and a depth of 9 hidden layers. Each neuron is represented by a real-valued, continuous quantity, described by the dynamical equation

$$\boldsymbol{S}(t+1) = g(\boldsymbol{S}(t) \cdot \boldsymbol{W}(\boldsymbol{t})),$$

where $\boldsymbol{S}(\boldsymbol{t}) \in \mathbb{R}^{\boldsymbol{N}}$ is the neuron state vector at time $t$, and $\boldsymbol{W}(\boldsymbol{t}) \in \mathbb{R}^{\boldsymbol{N}\times\boldsymbol{N}}$ the weighted adjacency matrix describing the coupling between the neurons. We use a scaled hyperbolic tangent activation function $g(x) = a\,tanh(b\,x)$ with $a = 1.7159$ and $b = 0.6666$, a configuration previously used for deep multilayer perceptrons[16–18]. The neurons between successive hidden layers are connected by all-to-all and asymmetric synaptic couplings, where the elements $\{w_{ij}\} \in \boldsymbol{W}$ are first drawn from a uniform distribution $[-0.05,0.05]$ and multiplied by a global scaling factor $m$, which serves as a control parameter for the effective coupling strength of the network. Varying $m$ moves the network between regimes in which activity is progressively attenuated or amplified as it propagates through successive layers.

### B. Measures of critical dynamics

To quantify this transition, we use two complementary dynamical measures: a branching parameter $\sigma$, describing the propagation of finite activity through the network[5,19], and the largest finite-time Lyapunov exponent $\lambda_0$, describing the maximal amplification of infinitesimal perturbations[20–23].

The calculation of $\sigma$ follows a simple concept: In the subcritical regime, when the connections between neurons are weak, signal activations decrease with each layer and $\sigma$<1. Conversely, in the supercritical regime, strong connections cause signal activations to increase with each layer and $\sigma$>1. At the critical point, activity propagates with $\sigma$=1, avoiding premature die-out or blow-up. To quantify this in DNNs, we define

$$\sigma = \frac{a_{\text{out}}}{a_{\text{in}}},$$

where $a_{\text{in}}$ and $a_{\text{out}}$ represent the average activities of the network's input (e.g. and image) and output (logits) layers, respectively. The activity of neuron $i$ in layer $l$ is calculated as the biased standard deviation of its activation across samples in a batch, $a_i^l = std_s(x_i^{l,s})$, where $s$ indexes the samples in a batch. The average activities in each layer are then calculated as

$$a_l = \frac{1}{N_l}\sum_{i=1}^{N_l} a_i^l,$$

where $N_l$ is the number of neurons in layer $l$.

The largest Lyapunov exponent of a dynamical system quantifies the average rate at which nearby trajectories diverge. We adapt this idea to DNNs by treating the layer-wise mapping as a discrete-time dynamical system and analyzing how small input perturbations grow or shrink as they propagate through it. This process is characterized by the finite-time Lyapunov exponent. A DNN with $N_0$ input components, $L$ hidden layers, $N_l$ neurons per hidden layer $l = 1, \dots, L$ and $N_{L+1}$ output neurons maps every input $\boldsymbol{x}^{(0)}$ to an output $\boldsymbol{x}^{(L+1)}$. The sensitivity of $x^{(l)}$ to small changes $\delta\boldsymbol{x}$ is determined by the linearization

$$\delta\boldsymbol{x}^{(l)} = \boldsymbol{D}^{(l)}\boldsymbol{W}^{(l)} \dots \boldsymbol{D}^{(2)}\boldsymbol{W}^{(2)}\boldsymbol{D}^{(1)}\boldsymbol{W}^{(1)}\delta\boldsymbol{x} = \boldsymbol{J}_l\delta\boldsymbol{x}.$$

Here, $W^{(l)}$ are the weight matrices in layer $l$, and $D^{(l)}$ are diagonal matrices with elements $D_{ij}^{(l)} = g'\left(b_i^{(l)}\right)\delta_{ij}$, where $b_i^{(l)} = \sum_{j=1}^{N_l} w_{ij}^{(l)} x_j^{(l-1)} - \Theta_i^{(l)}$ and $g'\left(b_i^{(l)}\right) = \frac{d}{db} g(b)|_{b=b_i^{(l)}}$ [20]. The function $g(\cdot)$ represents a non-linear activation function, and the weights $w_{ij}$ and thresholds $\Theta_i^{(l)}$ are parameters. The Jacobian matrix $\boldsymbol{J}_l(\boldsymbol{x})$ characterizes the growth or decay of small perturbations to $x$ [24]. Its maximal singular value $\Lambda_0^{(l)}$ increases or decreases exponentially as a function of $l$, with the rate $\lambda_0^{(l)} = l^{-1} \log\left(\Lambda_0^{(l)}(x)\right)$. The singular values $\Lambda_0^{(l)} > \Lambda_1^{(l)} > \dots > \Lambda_{N_l}^{(l)}$ are the square roots of the non-negative eigenvalues of the right Cauchy-Green tensor $\boldsymbol{J}_l^\top(\boldsymbol{x})\boldsymbol{J}_l(\boldsymbol{x})$. The maximal eigenvector of $\boldsymbol{J}_l^\top(\boldsymbol{x})\boldsymbol{J}_l(\boldsymbol{x})$ determines the direction of maximal stretching, i.e. in which input direction the output changes the most [20]. In practice, we calculate the Jacobian matrix using PyTorch's built-in automatic differentiation package.

## C. Local homeostatic adaptation

We next introduce a local adaptive rule designed to regulate neuronal activity without direct access to either of these global dynamical measures. The rule is inspired by adaptive-network models in which the activity of individual nodes provides a local signal for changes in network connectivity (Figure 1a).

We define the activity $a_i$ of a neuron $i$ based on how this neuron responds to different inputs. Specifically, we calculate the standard deviation of each neuron state across a batch of inputs (images) $a_i = std_s(x_i^s)$. A neuron that changes its state for different inputs has high activity, one that produces the same output regardless of the image has zero activity, respectively. We then set a threshold $T$, equal to the mean activity across the input layer (i.e. the images themselves) and classify each neuron as inactive if $a_i < T$ or active if $a_i \geq T$. After each forward pass (and optionally gradient-descent step), we then apply the following rule to a subset of randomly selected neurons $i$ of the network: If neuron $i$ does not change its state sufficiently across multiple inputs and is thus classified as inactive, a random subset of incoming links $w_{ij}$ is strengthened. If it changes its state significantly across inputs and is thus classified as active, a random subset of non-zero incoming links are weakened.

To be more specific, we describe a realization of this algorithm in detail: (i) Choose arbitrary starting condition for the deep neural network. (ii) Choose arbitrary inputs for the deep neural network (here we use the MNIST dataset). (iii) Optionally, do a forward pass though the network to apply gradient descent and do an optimizer step. (iv) During a second forward pass, in each layer $l$ a random subset $S_l$ of neurons is selected, and the activity $a_i$ of each neuron is determined as described above. (v) A random incoming link $w_{ij}$ is updated with

$$w'_{ij} = \begin{cases} (|w_{ij}| - c) \cdot sgn(w_{ij}) \text{ if } a_i > T \text{ (active)} \\ (|w_{ij}| + c) \cdot sgn(w_{ij}) \text{ if } a_i < T \text{ (inactive)} \end{cases}.$$

This results in a decreasing weight magnitude if the neuron is active, and an increasing weight magnitude if it is inactive. (vi) go to step (iii) and iterate. Starting from an arbitrary weight configuration, the adaptive dynamics therefore consist of repeated alternation between propagation of activity through the network and local modification of synaptic weights. When task learning is studied simultaneously, a standard gradient-descent update is performed first, followed by the local homeostatic update. This allows the task-dependent and homeostatic dynamics to act as two distinct adaptive processes on the same set of network parameters.

For numerical experiments involving supervised learning, we use the MNIST handwritten-digit dataset, comprising 60,000 training images and 10,000 testing images of handwritten digits (0-9), with each grayscale image sized $28 \times 28$ pixels. We train a simple Multi-Layer Perceptron (MLP) following the approach described in Ref.[25], implementing their largest configuration with 1,000 neurons in all hidden layers. Training uses standard backpropagation without momentum and a learning rate of $10^{-3}$. Input preprocessing consisted of mapping pixel intensities to real values in [−1.0, 1.0]. Exact network dimensions, learning rates, homeostatic step sizes and fractions of updated neurons or connections used in individual experiments are given in the corresponding figure captions.

## 3. Adaptive self-organization toward the critical point

We next asked whether the local homeostatic rule is sufficient to autonomously organize a deep network toward critical dynamics. We first characterized the underlying transition by varying the overall coupling strength of the network (Fig. 1). For weak coupling, activity progressively vanishes across layers, defining a subcritical regime; for strong coupling, activity is amplified, defining a supercritical regime. Between these regimes lies a critical point at which activity is propagated without systematic amplification or attenuation. We quantify this transition using two complementary measures: the branching parameter $\sigma$ which measures the average propagation of activity across the network, and the maximum Lyapunov exponent $\lambda_0 = 1/L \cdot \log(\Lambda_0)$ where $\Lambda_0$ is the maximum singular value of the input-output Jacobian. Criticality corresponds to $\sigma = 1$ and $\lambda_0 = 0$, separating vanishing from expanding activity and infinitesimal perturbations, respectively.

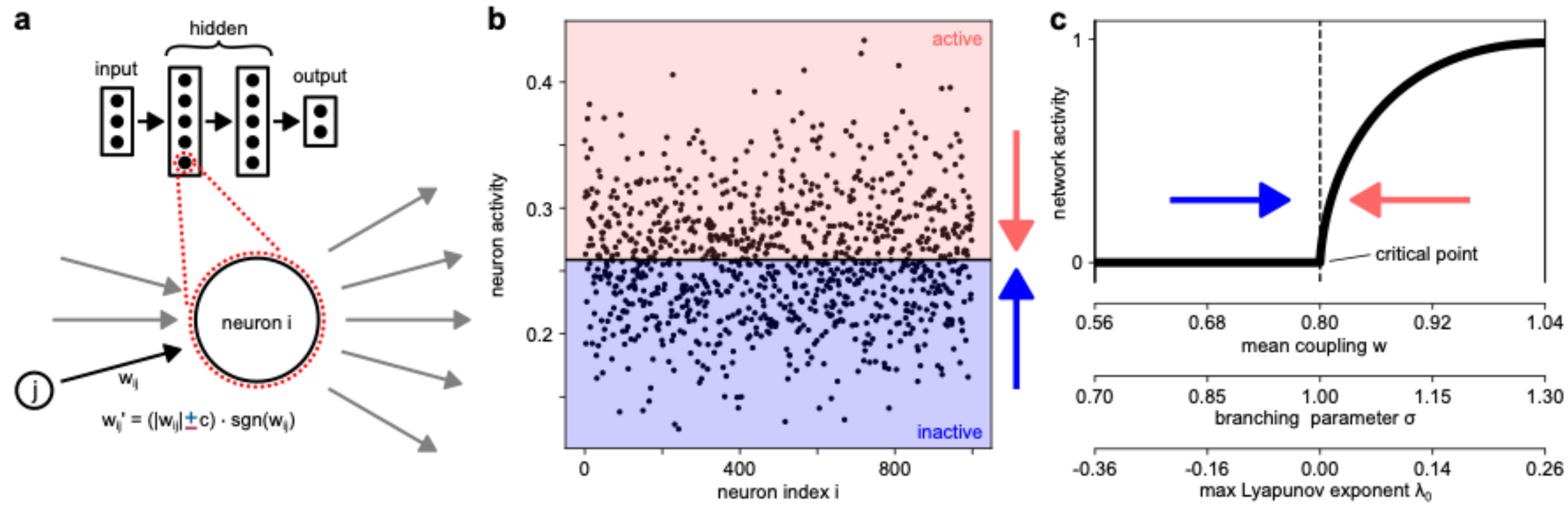

**Figure 1: A set of simple rules implements homeostatic plasticity in a deep neural network.** a, At each iteration, neuron $i$ modifies a randomly selected incoming connection: active neurons decrease its magnitude, whereas inactive neurons increase it. b, Activity $a_i$ of a neuron $i$ is defined by the standard deviation of activations relative to a threshold T (horizontal line). If $a_i < T$, the neuron is classified as inactive (blue region); if $a_i \geq T$, the neuron is classified as active (red region). Repeated local adaptation provides negative feedback on network activity (arrows). c, At the network level, weak coupling produces vanishing activity (subcritical dynamics), whereas strong coupling produces increasing activity (supercritical dynamics). The transition between these regimes defines the critical coupling strength and is characterized by $\sigma = 1$ and $\lambda_0 = 0$.

At the level of individual neurons, we define activity by the magnitude of the activation, $a_i$, relative to a threshold $T$ (Fig. 1). If $a_i \geq T$, the neuron is classified as active and decreases the magnitude of a randomly selected incoming connection by a fixed step $c$; if $a_i < T$, it is classified as inactive and increases the magnitude of a randomly selected incoming connection. Repeated application of this purely local rule drives individual neuronal activity toward the activity level imposed by the input distribution. Importantly, because increasing or decreasing incoming coupling also changes the propagation of activity through subsequent layers, this local homeostasis provides a negative-feedback mechanism on the global network state: subcritical networks strengthen their couplings, whereas supercritical networks weaken them.

To test whether this feedback robustly converges on the critical point, we initialized otherwise identical networks across a broad range of coupling strengths, spanning strongly subcritical to strongly supercritical dynamics, and iteratively applied the homeostatic rule during forward passes (Fig. 2). Despite their markedly different initial dynamics, all networks progressively converged toward the same stationary state. Activity became approximately conserved across network depth, the branching parameter converged to $\sigma = 1$, and the maximum Lyapunov exponent approached $\lambda_0 = 0$. At the same time, the initially distinct coupling strengths converged toward a common asymptotic value. Thus, simple neuron-local homeostatic adaptation robustly self-organizes deep networks toward a global critical point, without requiring explicit measurement or optimization of network-level criticality.

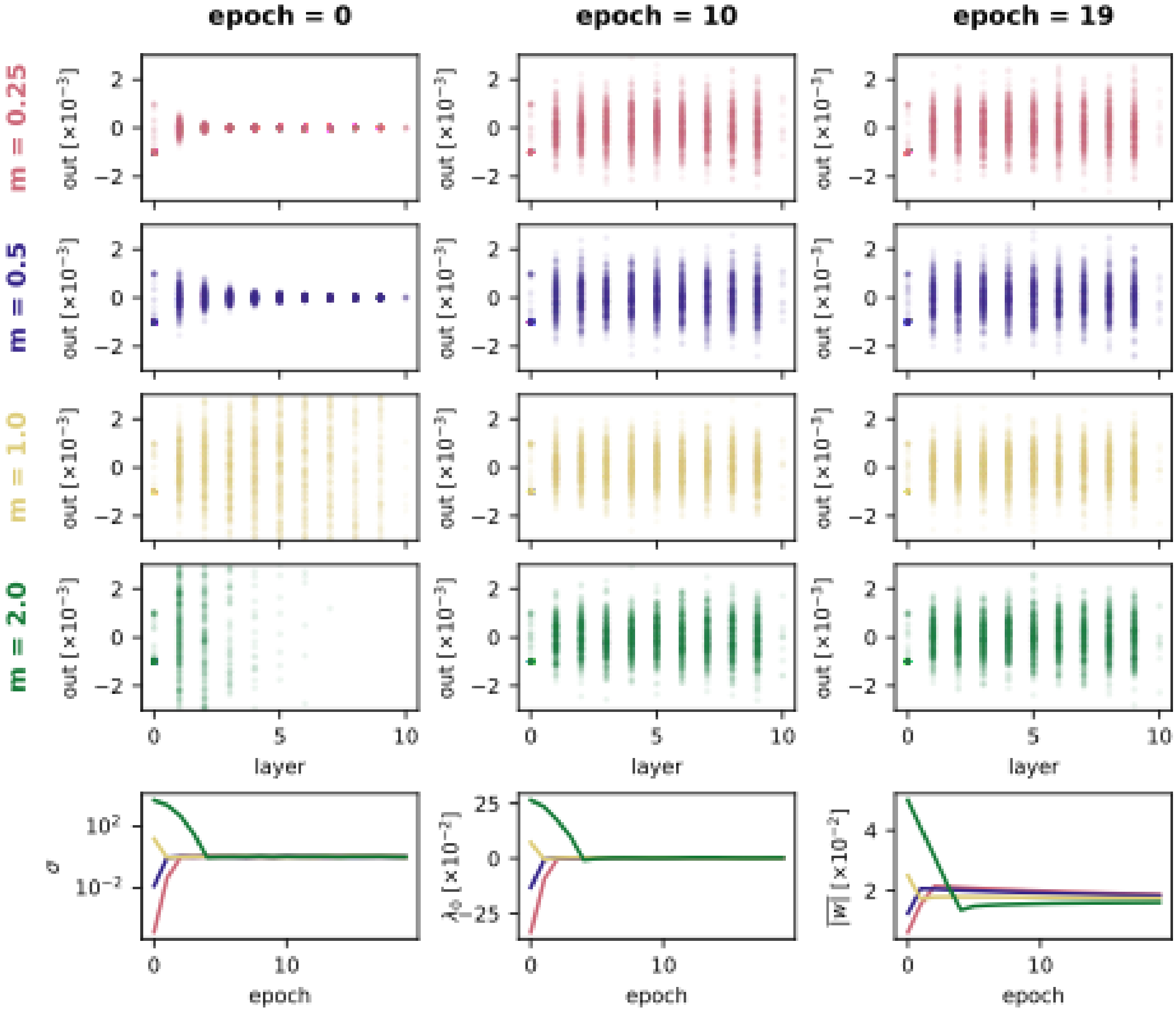

**Figure 2: Homeostatic plasticity autonomously drives deep neural networks to criticality**. Networks were initialized with uniform weights [–0.05, 0.05] and then scaled by a constant $m$ spanning subcritical ($m = 0.25$, neuron outputs vanishing within a few layers) to supercritical ($m = 2.0$, neuron outputs growing layer by layer) dynamics. During repeated homeostatic adaptation, initially vanishing or expanding activity converges toward stable propagation across network depth. Independent of initialization, networks approach $\sigma = 1, \lambda_0 = 0$, and a shared asymptotic average coupling strength $|\ \overline{w}\ |$. Networks contained nine hidden layers of 1,600 neurons with 784 input and 10 output neurons and were adapted for 20 epochs on MNIST images as inputs using $c = 0.001$ and batch size 32. Each iteration, 50% of neurons evaluated their activity and updated 1% of their incoming links. Lines show the mean of five runs, shaded regions denote ±1 standard deviation (narrower than line width); 0.2% of neurons are displayed for visual clarity, whereas all quantitative metrics were calculated using the full network.

## 4. Interaction with gradient-based learning

Thus far, we considered adaptive self-organized criticality (SOC) in isolation and used homeostatic plasticity as the sole mechanism to adapt network weights. While this allowed us to investigate whether the local plasticity rule can self-organize network dynamics towards criticality, these updates are task-agnostic: they do not optimize an objective function or use information about classification errors. In contrast, DNNs typically acquire task-specific representations through gradient-based learning. We therefore next examined how SOC interacts with conventional gradient descent when both learning mechanisms act simultaneously, considering both the resulting network dynamics and performance on the classification task.

To this end, we trained deep networks on MNIST using stochastic gradient descent (SGD), with each SGD update followed by an SOC update of varying step size (Fig. 3). In the absence of SOC, gradient-based learning progressively drove the network away from criticality and into the supercritical regime, as indicated by an increasingly positive maximum Lyapunov exponent (Fig. 3). Adding SOC counteracted this effect in a step-size-dependent manner: increasing the magnitude of the SOC updates progressively reduced the Lyapunov exponent, with sufficiently strong SOC maintaining the network close to the critical point throughout training. Thus, the self-organizing mechanism remained effective in the presence of gradient-based learning and could continuously oppose the dynamical changes induced by task optimization.

This stabilization of critical dynamics, however, came at a cost. Pure SGD resulted in the highest classification accuracy, whereas increasing the SOC step size progressively impaired task performance (Fig. 3). The same SOC updates that prevented gradient-based learning from driving the network into the supercritical regime therefore also interfered with optimization of the classification objective. To directly examine the relationship between the two update mechanisms, we calculated the cosine similarity between the SGD and SOC weight-update vectors (Fig. 3). Their low (negative) alignment indicates that SOC and SGD generally move the weights along substantially different directions in parameter space. This misalignment became progressively larger (more negative) with increasing SOC step size. Consequently, SOC, as implemented here, does not simply rescale or regularize the gradient trajectory but progressively redirects the network away from the parameter changes favoured by the task objective, thereby leading to an overall reduced performance. Together, these findings reveal a trade-off between dynamical criticality and task optimization when task-agnostic SOC and gradient-based learning act concurrently: stronger SOC more effectively preserves critical dynamics but increasingly interferes with the acquisition of task-specific performance. Aligning the directions of these two organizational principles may potentially solve this and lead to better performance, which will be the topic of future research.

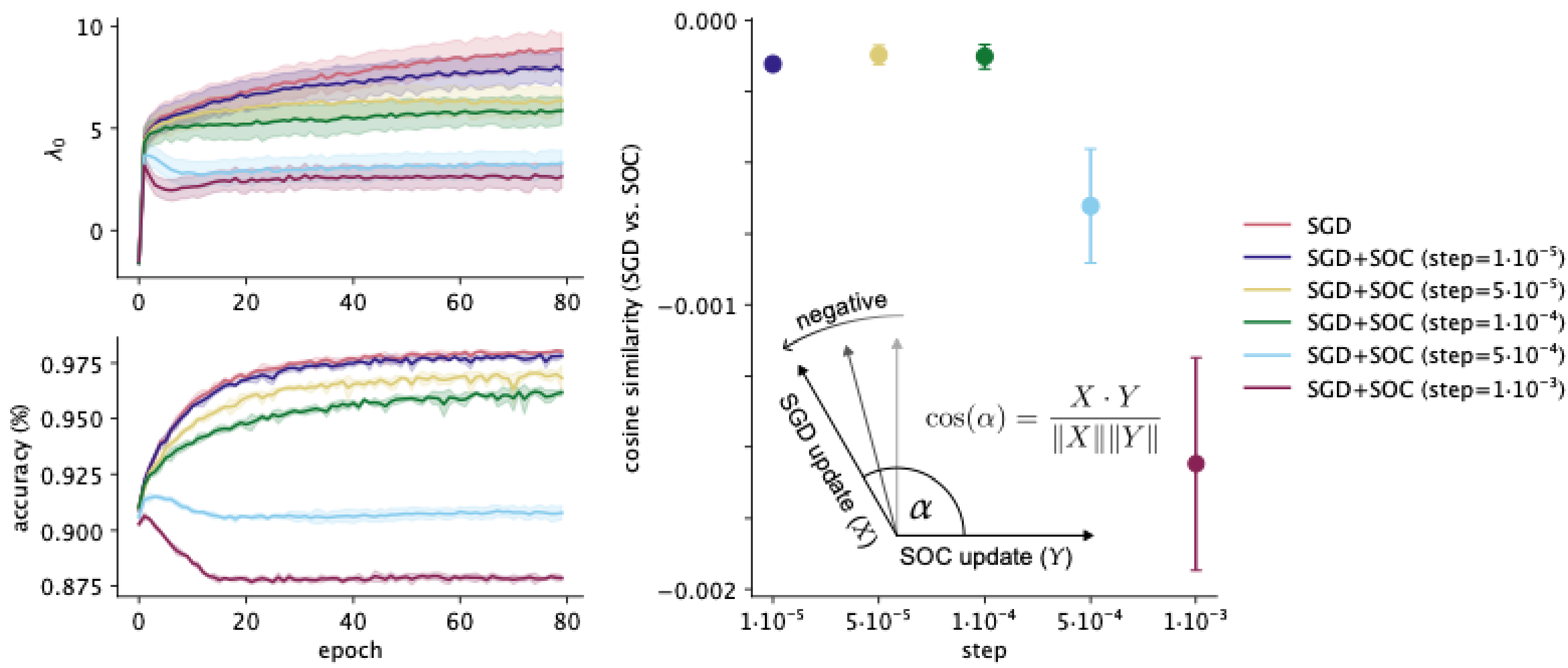

**Figure 3: Homeostatic plasticity–based self-organized criticality (SOC) promotes critical dynamics at the expense of task performance.** Deep neural networks were trained on MNIST using stochastic gradient descent (SGD) alone or SGD followed by an SOC update at each training iteration. Five SOC step sizes were tested (0.00001, dark blue; 0.00005, yellow; 0.0001, green; 0.0005, light blue; 0.001, lilac); pure SGD is shown in red. a, Maximum Lyapunov exponent, $\lambda_0$, over training epochs. SGD alone progressively drives the network into the supercritical regime ($\lambda_0 > 0$), whereas increasing the SOC step size increasingly counteracts this shift and maintains network dynamics closer to criticality ($\lambda_0 \approx 0$). b, Test classification accuracy for the same networks. Pure SGD achieves the highest accuracy, whereas task performance progressively decreases with increasing SOC step size. c, Cosine similarity between the SGD and SOC weight-update vectors for each SOC step size, averaged over the final training iterations of each epoch. The low cosine similarity indicates that SGD and SOC generally induce weight changes in substantially different directions in parameter space, providing a mechanistic explanation for the reduction in task performance observed in b. shaded regions and errorbars indicate ±1 standard deviation. Parameters: learning rate, 0.001; activation function, hyperbolic tangent; batch size, 32; weights initialized uniformly in (-0.005, 0.005); network architecture, nine hidden layers with 1,000 neurons each. In each SOC iteration, 50% of neurons were evaluated for their activity and 1% of their incoming links were updated.

## 5. Discussion

Our results show that DNNs can autonomously approach a critical dynamical regime through local, activity-dependent synaptic adaptation. This extends adaptive self-organization to criticality, previously demonstrated mainly in abstract threshold networks and models of biological neural systems, to modern DNNs. Importantly, no explicit information about the network-wide dynamical state is required: individual synapses are modified solely on the basis of local neuronal activity, yet the collective network dynamics converge towards the vicinity of the phase transition. Thus, in principle, self-organized criticality provides a mechanism by which DNNs could continuously regulate their dynamical operating point without explicitly computing a global criticality measure. The central mechanism is a negative feedback between local dynamics and the effective control parameter of the system. Local neuronal activity thereby acts as a sensor of the global dynamical phase: excessive activity induces weakening of synaptic interactions, whereas insufficient activity induces strengthening. Although the phase transition is a collective property of the entire network, local activity contains sufficient information about which side of the transition the system occupies to drive it back towards the critical region. We consider this local-to-global feedback to be the central physical mechanism underlying the observed self-organization.

The homeostatic rule used here also has a direct biological motivation. Activity-dependent homeostatic regulation of neuronal excitability and synaptic strength is well established experimentally[26,27], and related homeostatic mechanisms have previously been shown to regulate neuronal network models towards critical dynamics[11,28]. Our results therefore provide a conceptual bridge between these biological observations and artificial neural networks. From a machine-learning perspective, such autonomous regulation is particularly attractive because critical dynamics have been associated with favorable information-processing properties and may provide a means to stabilize learning and avoid dynamical failure modes. In our previous work, explicit regulation of criticality during gradient-based training improved robustness and prevented forms of model collapse[14].

In its present form, however, the SOC mechanism does not improve supervised learning. Rather, increasing the strength of homeostatic adaptation maintains the network closer to criticality while progressively reducing classification performance. The low alignment between gradient-based and SOC weight updates suggests a straightforward explanation: the two plasticity mechanisms optimize different objectives and therefore frequently move the network in different directions in parameter space. This competition is itself reminiscent of biological neural systems, in which multiple forms of synaptic plasticity coexist and must jointly regulate both function and stability. Here, task-dependent gradient plasticity and homeostatic plasticity similarly coexist, but are not yet coordinated. In our previous work, this problem could be overcome by incorporating the task objective and criticality objective into a common global loss function[14]. An analogous solution for locally self-organized learning may therefore require the task-dependent performance signal to enter the homeostatic update rule - for example through a global modulatory signal combined with local activity - so that criticality and task performance are optimized by a common plasticity mechanism. Developing such learning rules will be an important direction for future work.

Data availability